\documentstyle[12pt]{article}   
   
\newcommand{\ri}{\mbox{$\rm i$}}   
   
\newcommand{\rre}{\mbox{$\rm Re$}}   
   
\newcommand{\bfm}[1]{\mbox{\boldmath$#1$}}  
\newcommand{\ratio}[2]{\mbox{$#1\over#2$}}    
\newcommand{\tr}{\mbox{$\,\rm Tr$}}   
    
\begin{document}
\baselineskip=17pt
\parskip=5pt

\begin{titlepage} \footskip=5in     

\title{ 
\begin{flushright} \normalsize 
ISU-HET-99-15 \vspace{0.1em} \\ 
December 1999 \vspace{5em} \\  
\end{flushright}
\large\bf   
New Physics and Short-Distance  $\,s\rightarrow d\gamma\,$   
Transition in  $\,\Omega^-\rightarrow\Xi^-\gamma\,$  Decay  
}  

\author{\normalsize\bf 
Jusak~Tandean\thanks{E-mail: jtandean@iastate.edu} \\  
\normalsize\it 
Department of Physics and Astronomy, Iowa State University, 
Ames, IA 50011 
}
   
\date{}  
\maketitle  
   
\begin{abstract}  
We study the contributions of physics beyond the standard model 
to the short-distance  $\,s\rightarrow d\gamma\,$  transition 
in  $\,\Omega^-\rightarrow\Xi^-\gamma\,$  decay.  
We explore the possibility that the new interactions remove 
the chirality suppression which occurs in the standard-model 
contribution to  $\,s\rightarrow d\gamma\,$  and enhance the effect 
of this coupling by factors of a heavy-mass scale relative to 
the  $s$-quark mass.   
We consider two of the popular models for new physics, 
and, after taking into account constraints from other processes, 
find that their contributions can be larger than that of 
the standard model by up to a few times, which suggests that 
the yet unobserved  $\,\Omega^-\rightarrow\Xi^-\gamma\,$  decay 
is a likely probe for new physics. 
\end{abstract}   

\end{titlepage}

\section{Introduction}
    
Recent theoretical studies have shown that the decay  
$\,\Omega^-\rightarrow\Xi^-\gamma\,$  may be the leading candidate 
among weak radiative hyperon decays for providing a window 
to the short-distance  $\,s\rightarrow d\gamma\,$   
transition~\cite{singer}.  
The standard-model contribution to this short-distance transition 
in  $\,\Omega^-\rightarrow\Xi^-\gamma\,$  has been estimated to be 
smaller than long-distance contributions to the decay, but probably 
not significantly so~\cite{KogShi,EilIMS,NieBER}.   
In order to separate the various contributions to the decay, 
it may be necessary to study both its branching ratio and its 
asymmetry parameter~\cite{RafSin,EilIS}, but the only experimental 
information currently available is the upper limit~\cite{pdb}   
$\,{\rm BR}(\Omega^-\rightarrow\Xi^-\gamma)<4.6\times10^{-4}.\,$

In the absence of better data, one can take the position that   
the short-distance  $\,s\rightarrow d\gamma\,$  contribution   
to  $\,\Omega^-\rightarrow\Xi^-\gamma\,$  decay is not negligible 
compared to other contributions.  
In that case, the decay can also provide a window to new physics 
beyond the standard model, to the extent that the effects of new 
physics are not too small.

The short-distance  $\,s\rightarrow d\gamma\,$  transition is 
the second-generation analogue of the decay mode  
$\,b\rightarrow s\gamma,\,$  which has been used to place 
constraints on physics beyond the standard model~\cite{bsg1}.  
The mode is particularly useful for constraining new interactions 
which remove the chirality suppression that occurs in the standard 
model.  
In this case, the amplitude is enhanced by factors of a heavy-mass  
scale relative to the  $b$-quark mass.   
The same type of new physics enhances the  $\,s\rightarrow d\gamma\,$   
transition by a factor of a heavy-mass scale relative to  
the $s$-quark mass. 
In models in which the enhancement is as large as one can expect on 
dimensional grounds, it is possible to place interesting constraints 
from  $\,s\rightarrow d\gamma\,$  on the new interactions,  
especially when the decay modes involved are not significantly 
dominated by long-distance physics.

In this paper, we would like to explore this possibility in  
$\,\Omega^-\rightarrow\Xi^-\gamma\,$  decay  and consider two 
types of models in which the short-distance transition can be 
significantly enhanced with respect to the standard model (SM).   
This study will be complementary to studies on 
new-physics contributions to  $\,s\rightarrow d\gamma\,$  in kaon 
and other hyperon  decays~\cite{dsg}.      
In Ref.~\cite{RosBEN}, the impact of new interactions on  
$\,s\rightarrow d\gamma\,$  in  $\,\Omega^-\rightarrow\Xi^-\gamma\,$ 
was also investigated, but the models considered there did not 
remove the chirality suppression occurring in the SM, resulting 
in negligible new contributions.

In the next section, we introduce our notation for the effective 
interaction responsible for the short-distance  
$\,s\rightarrow d\gamma\,$  transition.   
In order to describe the effect of this transition on  
$\,\Omega^-\rightarrow\Xi^-\gamma\,$  decay, 
we shall use a chiral-Lagrangian analysis.   
In the following two sections, we consider contributions of 
new physics to  $\,s\rightarrow d\gamma\,$  in 
left-right symmetric models~\cite{mohapatra,FujYam,bsg2}  and 
in generic supersymmetric models~\cite{GabGMS}.  
We are not here interested in the specific details of the models, 
and so we will consider only the effective low-energy operators that 
the models may generate.  
The last section contains our conclusions.

\section{Short-distance  $\,s\rightarrow d\gamma\,$  operator
in  $\,\Omega^-\rightarrow\Xi^-\gamma\,$  decay}
    
The effective Hamiltonian responsible for the short-distance  
$\,s\rightarrow d\gamma,d g\,$  transitions can be written, 
following the notation of  Ref.~\cite{BurCIRS},  as  
\begin{eqnarray}   \label{H_eff}
{\cal H}_{\rm eff}^{}  \;=\;  
C_{\gamma}^+ Q_{\gamma}^+ + C_{\gamma}^- Q_{\gamma}^- 
+ C_{g}^+ Q_{g}^+ + C_{g}^- Q_{g}^- 
\;+\;  {\rm h.c.}   \;,    
\end{eqnarray}      
where  $C_{\gamma,g}^\pm$  are the Wilson coefficients and  
\begin{eqnarray}   
\begin{array}{c}   \displaystyle  
Q_{\gamma}^\pm  \;=\;    
{e Q_d^{}\over16\pi^2} \bigl( 
\bar{d}_{\rm L}^{}\, \sigma_{\mu\nu}^{}\, s_{\rm R}^{}  
\pm \bar{d}_{\rm R}^{}\, \sigma_{\mu\nu}^{}\, s_{\rm L}^{}  
\bigr) F^{\mu\nu}   \;,    
\vspace{1ex} \\   \displaystyle  
Q_{g}^\pm  \;=\;    
{g_s^{}\over16\pi^2} \bigl( 
\bar{d}_{\rm L}^{}\, \sigma^{\mu\nu} T_a^{}\, s_{\rm R}^{}  
\pm \bar{d}_{\rm R}^{}\, \sigma^{\mu\nu} T_a^{}\, s_{\rm L}^{}  
\bigr) G_a^{\mu\nu} 
\end{array}      
\end{eqnarray}      
are the so-called electromagnetic- and chromomagnetic-dipole 
operators, respectively, with  $\,eQ_d^{}=-e/3\,$  being 
the  $d$-quark  charge and  $F^{\mu\nu}$  
$ \bigl( G_a^{\mu\nu} \bigr) $  being the photon (gluon)  
field-strength tensor.  
We notice that  $Q_{\gamma}^\pm$  transform as  
$\,\bigl(\bar{3}_{\rm L}^{},3_{\rm R}^{}\bigr)\pm  
\bigl(3_{\rm L}^{},\bar{3}_{\rm R}^{}\bigr)$  under chiral rotations, 
and so  $Q_{\gamma}^{+}$  $ \bigl( Q_{\gamma}^{-} \bigr) $ 
is even (odd) under parity.  
$C_{\gamma,g}^\pm$  contain standard-model and possible new-physics 
contributions.

To evaluate the contribution of the  $\,s\rightarrow d\gamma\,$  
operators in  Eq.~(\ref{H_eff})  to  
$\,\Omega^-\rightarrow\Xi^-\gamma\,$  decay,  one needs 
to calculate  $\,\bar{d}\sigma_{\mu\nu}(1\pm\gamma_5^{})s\,$  
being sandwiched between hadronic states.  
Unfortunately, there is presently no reliable way to compute 
these matrix elements.  
For this reason, we will estimate the impact of the operators by   
employing a chiral-Lagrangian approach combined with naive   
dimensional-analysis~\cite{nda}.

The chiral Lagrangian that describes  
$\,\Omega^-\rightarrow\Xi^-\gamma\,$  is written down in terms of   
a  $3\times3$ matrix  $B$  containing the baryon-octet fields, 
a Rarita-Schwinger tensor  $T_{abc}^\mu$  representing   
the decuplet baryons, and field-strength tensors  $\ell_{\mu\nu}^{}$  
and  $r_{\mu\nu}^{}$,  where  
$\,\ell_{\mu\nu}^{}=r_{\mu\nu}^{}=eQF_{\mu\nu}^{},\,$  with   
$\,Q={\rm diag}(2,-1,-1)/3\,$  being the quark-charge matrix.  
Also included is the exponential 
$\,\xi=\exp \bigl[ \ri\phi/ \bigl( 2f_{\!\pi}^{} \bigr) \bigr] ,\,$  
where  $\phi$  is the usual $3\times3$ matrix  containing the octet 
of pseudo-Goldstone bosons  and  
$\,f_{\!\pi}^{}\approx92.4\,\rm MeV\,$  is the pion-decay constant.  
(The notation here for hadronic fields follows that of 
Ref.~\cite{AbdTV}.)   
Under chiral rotations,  $\,B\rightarrow U B U^\dagger,\,$  
$\,T_{abc}^\mu\rightarrow U_{ad}^{} U_{be}^{} U_{cf}^{}
T_{def}^\mu,\,$  
$\,\ell_{\mu\nu}^{}\rightarrow L \ell_{\mu\nu}^{} L^\dagger,\,$  
and  $\,r_{\mu\nu}^{}\rightarrow R r_{\mu\nu}^{} R^\dagger,\,$  
where  $\,L,R\in\rm SU(3)_{L,R}^{}\,$   
and  $U$  is implicitly defined by the transformation
$\,\xi\rightarrow L\xi U^\dagger=U\xi R^\dagger.\,$

There are many possible chiral realizations of 
the  $\,s\rightarrow d\gamma\,$  operators in  
${\cal H}_{\rm eff}^{}$.    
As an example at lowest order, we write down 
\begin{eqnarray}   \label{L_(TBg)}  
{\cal L}_{TB\gamma}^{}  \;=\;  
\epsilon_{ace}^{}\, \bar{B}_{ab}^{}\, 
\Bigl( \xi^\dagger \ell_{\mu\nu}^{} h \xi^\dagger  
      + \xi r_{\mu\nu}^{} h \xi \Bigr) _{cd}^{}\, 
\bigl( \alpha_-^{}+\alpha_+^{}\gamma_5^{} \bigr) 
\bigl( \gamma^\mu T_{bde}^\nu-\gamma^\nu T_{bde}^\mu \bigr)  
\;+\;  {\rm h.c.}   \;,      
\end{eqnarray}      
where  $\,h= \bigl( \lambda_6^{}+\ri\lambda_7^{} \bigr) /2\,$  
selects out  $\,s\rightarrow d\gamma\,$  transitions, the terms 
proportional to the constants  $\alpha_\pm^{}$  have 
the chiral-transformation properties of  $Q_\gamma^\pm$,  
and we have included only terms contributing to  
$\,\Omega^-\rightarrow\Xi^-\gamma.\,$  
Using naive dimensional-analysis~\cite{nda}, we obtain 
the order-of-magnitude estimate   
\begin{eqnarray}   
\alpha_\pm^{}  \;=\;  {C_\gamma^\pm\over16\pi^2}   \;.    
\end{eqnarray}      

The most general, gauge-invariant form of the amplitude for  
$\,\Omega^-(p_\Omega^{})\rightarrow \Xi^-(p_\Xi^{})\gamma(q)\,$  
can be written as~\cite{RafSin}     
\begin{eqnarray}   
{\cal M}  \;=\;  
\ri e G_{\rm F}^{}\, \bar{u}^{(\Xi)}\, \Biggl[   
\bigl( a_1^{}+a_2^{}\gamma_5^{} \bigr)   
\bigl( \not{\!q} g^{\mu\nu}-\gamma^\mu q^\nu \bigr)     
+ \bigl( b_1^{}+b_2^{}\gamma_5^{} \bigr)   
 { p_\Omega^{}\cdot q\, g^{\mu\nu}-p_\Omega^\mu q^\nu  \over  
  m_\Omega^{} } 
\Biggr] \, \varepsilon_\mu^{*} u_\nu^{(\Omega)}   \;,   
\end{eqnarray}      
where  $G_{\rm F}^{}$  is the Fermi constant, the  $u$'s  are 
baryon spinors, and  $a_{1,2}^{},b_{1,2}^{}$  are constants to be 
determined from the Lagrangian.  
The corresponding decay width is given by~\cite{RafSin} 
\begin{eqnarray}   
\Gamma_{\Omega\rightarrow\Xi\gamma}^{}  \;=\;  
\ratio{2}{3}\alpha G_{\rm F}^2\, |\bfm{q}|^3 
\begin{array}[t]{l}   \displaystyle   
\Biggl[ 
\Biggl( 1+{p_\Omega^{}\cdot p_\Xi^{}\over m_\Omega^2} \Biggr)  
\bigl( a_1^2+a_2^2 + a_1^{} b_1^{}+a_2^{} b_2^{} \bigr)  
+ {p_\Omega^{}\cdot p_\Xi^{}\over m_\Omega^2}  
 \bigl( b_1^2+b_2^2 \bigr)
\vspace{2ex} \\   \displaystyle   
\;\;+\; 
{m_\Xi^{}\over m_\Omega^{}} 
\bigl( 2a_1^{} b_1^{}-2a_2^{} b_2^{} + b_1^2-b_2^2 \bigr) \;   
\Biggr]   \;.     
\end{array}      
\end{eqnarray}      
Then, from the Lagrangian in Eq.~(\ref{L_(TBg)}),  we derive    
\begin{eqnarray}   
a_1^{}  \;=\;  {C_\gamma^-\over 12\pi^2 G_{\rm F}^{}}   \;,  
\hspace{3em}  
a_2^{}  \;=\;  {C_\gamma^+\over 12\pi^2 G_{\rm F}^{}}   \;,  
\hspace{3em}  
b_1^{}  \;=\;  b_2^{}  \;=\;  0   \;,       
\end{eqnarray}      
leading to a branching ratio 
\begin{eqnarray}   \label{br}
{\rm BR} \bigl( \Omega^-\rightarrow\Xi^-\gamma \bigr)  
\;\approx\;  
2.4\times 10^{6}\;  
\Bigl[ \bigl( C_\gamma^+ \bigr) ^2 
      + \bigl( C_\gamma^- \bigr) ^2 \Bigr] \; \rm GeV^2   \;.   
\end{eqnarray}      

In the SM,  the Wilson coefficients at the one-loop level 
without QCD corrections are given  by~\cite{InaLim,BerFG}   
\begin{eqnarray}   
\begin{array}{c}   \displaystyle     
C_{\gamma,\rm SM}^{\pm} (m_W^{}) \;=\;   
{G_{\rm F}^{}\over\sqrt{2}}\, (m_s^{}\pm m_d^{})\,  
\sum_{q=u,c,t} V_{qd}^{*} V_{qs}^{}\,   
{F_{\rm SM}^{}(x_q^{})\over Q_d^{}}   \;,  
\vspace{2ex} \\   \displaystyle   
C_{g,\rm SM}^{\pm} (m_W^{})  \;=\;   
{G_{\rm F}^{}\over\sqrt{2}}\, (m_s^{}\pm m_d^{})\,  
\sum_{q=u,c,t} V_{qd}^{*} V_{qs}^{}\, G_{\rm SM}^{}(x_q^{})   \;,  
\end{array}      
\end{eqnarray}      
where  $\,x_q^{}=m_q^2/m_W^2\,$  and     
\begin{eqnarray}   
\begin{array}{c}   \displaystyle     
F_{\rm SM}^{} (x)  \;=\;  
-{3x^3-2x^2\over 2 (x^{}-1)^4}\, \ln x^{}   
+ {8x^3+5x^2-7x\over 12 (x-1)^3}   \;,  
\vspace{2ex} \\   \displaystyle   
G_{\rm SM}^{} (x)  \;=\;  
{3x^2\over 2(x-1)^4}\, \ln x^{}   
+ {x^3-5x^2+2x\over 4(x-1)^3}   \;.   
\end{array}      
\end{eqnarray}      
Although  $C_{\gamma,\rm SM}^{\pm}(\mu)$  are very small at  
$\,\mu=m_W^{},\,$  they receive large QCD corrections at 
$\,\mu\sim 1\,\rm GeV,\,$  mostly due to the mixing of 
$Q_{\gamma}^{\pm}$  with the four-quark operator  
$\,Q_2^{}=\bar{d}_{\rm L}^{}\gamma^\mu u_{\rm L}^{}
\bar{u}_{\rm L}^{}\gamma_\mu^{}
s_{\rm L}^{}\,$~\cite{BerFG,ShiVZ,BucBL}.  
For our numerical estimates, we will use  
$\,\alpha_s^{}(m_Z^{})=0.119,\,$  the middle values of the quark 
masses in Ref.~\cite{pdb},  and  the CKM-matrix elements in 
the Wolfenstein parameterization with  $\,\lambda =0.22$,   
$\,A=0.82$,  $\,\rho=0.16\,$  and  $\,\eta=0.38\,$  
from Ref.~\cite{mele}.  
At  $\,\mu=1\,\rm GeV,\,$  we find    
\begin{eqnarray}   \label{csm}  
C_{\gamma,\rm SM}^+(\mu)  \;\approx\;   
-5.0\times 10^{-7}\; {\rm GeV}^{-1}   \;,     
\hspace{3em}  
C_{\gamma,\rm SM}^-(\mu)  \;\approx\;   
-4.5\times 10^{-7}\; {\rm GeV}^{-1}   \;,     
\end{eqnarray}      
where we have neglected the small imaginary parts.     
Employing these numbers in Eq.~(\ref{br}),  we obtain for 
the short-distance contribution  
\begin{eqnarray}   
{\rm BR} \bigl( \Omega^-\rightarrow\Xi^-\gamma \bigr) _{\rm SM}^{}   
\;\approx\;  1.1\times 10^{-6}   \;, 
\end{eqnarray}      
which is within the range of previous 
estimates using other methods~\cite{EilIMS,NieBER}.

\section{Left-right symmetric models\label{RHW}}
   
One of the popular extensions of the SM is the left-right symmetric 
model~\cite{mohapatra}, which incorporates the mixing of left- and 
right-handed  $W$ bosons.    
Variations of this model have been studied in the context of  
$\,b\rightarrow s\gamma\,$  in detail~\cite{FujYam,bsg2}, 
where an enhancement by a factor of  $m_t^{}/m_b^{}$  occurring in 
this process has led to constraints on the right-handed  $tbW$  
coupling. 
Similar interactions can occur in  $\,s\rightarrow d\gamma,\,$   
which replace the suppression due to light-quark mass in  
$C_{\rm SM}^{}$  with an enhancement from a heavy-quark mass.

For our purposes, it suffices to deal with the effective Lagrangian   
that results after integrating out the heavy right-handed $W$   
at the scale of the standard-$W$ mass.    
This can be done easily by following the formalism of  
Ref.~\cite{PecZha}.   
In the unitary gauge, the new interaction can be written as 
\begin{eqnarray}   
{\cal L}_{\rm RH}^{}  \;=\;  
-{g_2^{}\over\sqrt{2}} 
\Bigl( \bar{u}_{\rm R}^{} \hspace{1em} \bar{c}_{\rm R}^{} 
      \hspace{1em} \bar{t}_{\rm R}^{} \Bigr) \, \gamma^\mu\, 
\tilde{V} 
\left( \begin{array}{c}   \displaystyle  
d_{\rm R}^{}   \vspace{1ex} \\   \displaystyle   
s_{\rm R}^{}   \vspace{1ex} \\   \displaystyle  
b_{\rm R}^{}  
\end{array} \right) 
W_\mu^+    
\;+\;  {\rm h.c.}   \;,  
\end{eqnarray}      
where  $\tilde{V}$  is a $3\times3$  unitary matrix  having elements 
$\,\tilde{V}_{qq'}^{}=V_{qq'}^{}\kappa_{qq'}^{\rm R},\,$  
with  $V_{qq'}^{}$  being  CKM-matrix elements and  
$\kappa_{qq'}^{\rm R}$  real numbers.   
In writing  ${\cal L}_{\rm RH}^{}$  above, we have assumed  $C\!P$  
invariance and neglected modifications to the left-handed 
$W$-couplings  which do not lead to enhanced effects.

The interaction  ${\cal L}_{\rm RH}^{}$  has previously been   
considered in a  $\,b\rightarrow s\gamma\,$  study~\cite{FujYam}  
and in a study of  $C\!P$  violation in  $B$  decays~\cite{AbdVal}.      
Generalizing those results, we can derive the Wilson coefficients 
\begin{eqnarray}     
\begin{array}{c}   \displaystyle   
C_{\gamma,\rm RH}^{\pm} (m_W^{})  \;=\;   
{G_{\rm F}^{}\over\sqrt{2}}   
\sum_{q=u,c,t} V_{qd}^{*} V_{qs}^{}\,   
\Bigl( \kappa_{qs}^{\rm R}\pm\kappa_{qd}^{\rm R*} \Bigr) \, m_q^{}\,    
{F_{\rm RH}^{}(x_q^{})\over Q_d^{}}   \;,   
\vspace{2ex} \\   \displaystyle   
C_{g,\rm RH}^{\pm} (m_W^{})  \;=\;   
{G_{\rm F}^{}\over\sqrt{2}}   
\sum_{q=u,c,t} V_{qd}^{*} V_{qs}^{}\,     
\Bigl( \kappa_{qs}^{\rm R}\pm\kappa_{qd}^{\rm R*} \Bigr) \, m_q^{}\,  
G_{\rm RH}^{}(x_q^{})   \;,  
\end{array}      
\end{eqnarray}      
where  
\begin{eqnarray}   
F_{\rm RH}^{} (x)  \;=\;   
{-3x^2+2x\over(x-1)^3} \, \ln x - {5x^2-31x+20\over 6(x-1)^2}   \;,    
\hspace{2em}   
G_{\rm RH}^{} (x)  \;=\;   
{6x\, \ln x\over(x-1)^3} - {3+3x\over (x-1)^2} - 1   \;.    
\end{eqnarray}      
At a hadronic-mass scale  $\,\mu\sim 1\,\rm GeV,\,$  
the coefficients of the electromagnetic-dipole operators 
generated by the new couplings are~\cite{running}  
\begin{eqnarray}   \label{running}
C_{\gamma,\rm RH}^{\pm} (\mu)  \;=\;   
\eta^{16/(33-2n_{\rm f}^{})}\, C_{\gamma,\rm RH}^\pm(m_W^{})  
+ 8 \Bigl[ \eta^{16/(33-2n_{\rm f}^{})} 
          - \eta^{14/(33-2n_{\rm f}^{})} \Bigr]  
 C_{g,\rm RH}^\pm(m_W^{})   \;,  
\end{eqnarray}      
where  $\,\eta=\alpha_{\rm s}^{}(m_W^{})/\alpha_{\rm s}^{}(\mu)\,$     
and  $n_{\rm f}^{}$  is the number of active quarks.   
We then have at  $\,\mu=1\,\rm GeV\,$  
\begin{eqnarray}     
C_{\gamma,\rm RH}^{\pm}(\mu)  \;\approx\; 
\Bigl[    
\Bigl( \kappa_{us}^{\rm R}\pm\kappa_{ud}^{\rm R*} \Bigr) 
- 243\, \Bigl( \kappa_{cs}^{\rm R}\pm\kappa_{cd}^{\rm R*} \Bigr)   
- (20 + 9\, \ri) 
 \Bigl( \kappa_{ts}^{\rm R}\pm\kappa_{td}^{\rm R*} \Bigr)   
\Bigr] \times 10^{-7} \;{\rm GeV}^{-1}   \;.  
\end{eqnarray}      

If all the  $\kappa^{\rm R}$  parameters  are of the same order 
of magnitude, then the $c$-quark contribution will be dominant.   
It has been found in  Ref.~\cite{FujYam} that  
$\,b\rightarrow s\gamma\,$  
constrains  $\kappa_{tb}^{\rm R}$  to be at most a 
few percent.  
We can, therefore, reasonably assume that other  
$\kappa^{\rm R}$'s  have similar upper-bounds.   
Thus, taking  $\,\kappa_{cs}^{\rm R}=0.02\,$  and  
$\,\kappa_{cd}^{\rm R}=0,\,$  and neglecting contributions from  
$u$  and  $t$  quarks,  we find   
\begin{eqnarray}     
C_{\gamma,\rm RH}^{\pm}(\mu)  \;\approx\;  
- 4.9\times 10^{-7} \;{\rm GeV}^{-1}   \;,    
\end{eqnarray}      
which is comparable to the SM values in Eq.~(\ref{csm}).    
This order-of-magnitude estimate of  $C_{\gamma,\rm RH}^{\pm}$  
suggests that it is possible for left-right symmetric models to 
generate contributions to  $\,s\rightarrow d\gamma\,$  which 
exceed that of the SM.

\section{Supersymmetric models\label{susy}}   
  
Another popular scenario for new physics in which the coefficients  
$C_{\gamma,g}^{}$  can be naturally large is the general 
supersymmetric extension of the SM.  
In this class of models, one can generate 
the  $\,s\rightarrow d\gamma\,$  operators at one-loop 
via intermediate squarks and gluinos. 
The enhancement is due both to the strong coupling constant and 
to the removal of chirality suppression that results in 
an enhancement factor of a gluino mass relative to 
a light-quark mass.  
In order to avoid specific models, we follow  Ref.~\cite{GabGMS} 
to work in the so-called mass-insertion approximation.

If SUSY particles are integrated out at the scale  
$\,m_{\tilde{g}}^{}>m_t^{},\,$  where  $m_{\tilde{g}}^{}$  is 
the average gluino-mass, then at  $\,\mu=1\,\rm GeV\,$  the Wilson 
coefficients arising from  the new physics 
are\footnote{Here, we use the expression given in  
Ref.~\cite{BurCIRS},  instead of  Eq.~(\ref{running}).}      
\begin{eqnarray}   
C_{\gamma,\rm susy}^\pm (\mu)  \;=\;    
\bar{\eta}^2\,  
C_{\gamma,\rm susy}^\pm \bigl( m_{\tilde{g}}^{} \bigr)  
+ 8 \bigl( \bar{\eta}^2-\bar{\eta} \bigr) \, 
 C_{g,\rm susy}^\pm \bigl( m_{\tilde{g}}^{} \bigr)   \;,  
\end{eqnarray}      
where  
\begin{eqnarray}   
\bar{\eta}  \;=\;    
\Biggl( 
{ \alpha_{\rm s}^{} \bigl( m_{\tilde{g}}^{} \bigr)  \over  
 \alpha_{\rm s}^{} \bigl( m_t^{} \bigr) } \Biggr) ^{2\over21} \,   
\Biggl( 
{ \alpha_{\rm s}^{} \bigl( m_t^{} \bigr)  \over  
 \alpha_{\rm s}^{} \bigl( m_b^{} \bigr) } \Biggr) ^{2\over23} \,   
\Biggl( 
{ \alpha_{\rm s}^{} \bigl( m_b^{} \bigr)  \over  
 \alpha_{\rm s}^{} \bigl( m_c^{} \bigr) } \Biggr) ^{2\over25} \, 
\Biggl( 
{ \alpha_{\rm s}^{} \bigl( m_c^{} \bigr)  \over  
 \alpha_{\rm s}^{} \bigl( \mu \bigr) } \Biggr) ^{2\over27}   \;.  
\end{eqnarray}      
The full expressions for  
$C_{\rm susy}^{\pm} \bigl( m_{\tilde{g}}^{} \bigr) $  can be 
found  in  Ref.~\cite{GabGMS}.    
We are interested here only in the terms proportional to  
$m_{\tilde{g}}^{}/m_s^{}$,  which may give potentially large 
contributions to  $\,s d(\gamma,g)\,$  couplings  and  are given 
by~\cite{BurCIRS}    
\begin{eqnarray}     
\begin{array}{c}   \displaystyle   
C_{\gamma,\rm susy}^{\pm} (m_{\tilde{g}}^{})  \;=\;   
{\pi\, \alpha_{\rm s}^{}(m_{\tilde{g}}^{})\over m_{\tilde{g}}^{}}
\Bigl[ \bigl( \delta_{12}^{d} \bigr) _{\rm LR}^{} 
      \pm \bigl( \delta_{12}^{d} \bigr) _{\rm RL}^{} \Bigr] 
F_{\rm susy}^{}(x_{gq}^{})   \;,  
\vspace{2ex} \\   \displaystyle   
C_{g,\rm susy}^{\pm} (m_{\tilde{g}}^{})  \;=\;   
{\pi\, \alpha_{\rm s}^{}(m_{\tilde{g}}^{})\over m_{\tilde{g}}^{}}
\Bigl[ \bigl( \delta_{12}^{d} \bigr) _{\rm LR}^{} 
      \pm \bigl( \delta_{12}^{d} \bigr) _{\rm RL}^{} \Bigr] 
G_{\rm susy}^{}(x_{gq}^{})   \;,  
\end{array}      
\end{eqnarray}      
where the  $\delta$'s  are the parameters of the mass-insertion 
formalism,  $\,x_{gq}^{}=m_{\tilde{g}}^2/m_{\tilde{q}}^2,\,$  
with  $m_{\tilde{q}}$  being the average squark-mass,  and  
\begin{eqnarray}   
\begin{array}{c}   \displaystyle   
F_{\rm susy}^{} (x)  \;=\;   
{4x\, \bigl( 1+4x-5x^2+4x\,\ln x+2x^2\,\ln x \bigr) \over 3(1-x)^4}   \;,    
\vspace{2ex} \\   \displaystyle   
G_{\rm susy}^{} (x)  \;=\;   
{ x\, \bigl( 22-20x-2x^2-x^2\,\ln x+16x\,\ln x+9\,\ln x \bigr)  \over  
 3(1-x)^4 }   \;.    
\end{array}      
\end{eqnarray}      
Numerically, we obtain, for  
$\,m_{\tilde{q}}^{}=500\,\rm GeV,\,$    
\begin{eqnarray}     
{ C_{\gamma,\rm susy}^{\pm} (1\,{\rm GeV})  \over  
 \Bigl[ \bigl( \delta_{12}^{d} \bigr) _{\rm LR}^{} 
       \pm \bigl( \delta_{12}^{d} \bigr) _{\rm RL}^{} \Bigr] }
\;\approx\;   
\left\{  
\begin{array}{ll}   \displaystyle   
4.3\times 10^{-4}\;{\rm GeV}^{-1}   
&  {\rm for}\; x_{gq}^{}  \;=\;  0.3 
\vspace{1ex} \\   \displaystyle   
2.5\times 10^{-4}\; {\rm GeV}^{-1}   
&  {\rm for}\; x_{gq}^{}  \;=\;  1 
\vspace{1ex} \\   \displaystyle   
1.0\times 10^{-4}\;{\rm GeV}^{-1}   
&  {\rm for}\; x_{gq}^{}  \;=\;  4 
\end{array}      
\right.  
\;.   
\end{eqnarray}      
It has been shown in  Refs.~\cite{GabGMS,MasMur} 
that the real part of  $\bigl( \delta_{12}^{d} \bigr) _{\rm LR}^{}$  
is constrained from  $\Delta m_{K}^{}$  to be typically less than 
$10^{-2}$,  
and that its imaginary part is expected from  $\epsilon'/\epsilon$   
to be less than a few times  $10^{-5}$.   
Using the upper bounds of  
$\rre \bigl( \delta_{12}^{d} \bigr) _{\rm LR}^{} $  
from  Ref.~\cite{MasMur} and choosing  
$\, \bigl( \delta_{12}^{d} \bigr) _{\rm RL}^{}=0,\,$  
we find   
\begin{eqnarray}   \label{C,susy}   
C_{\gamma,\rm susy}^{\pm} (1\,{\rm GeV})  \;\approx\;   
\left\{  
\begin{array}{ll}   \displaystyle   
33\times 10^{-7}\;{\rm GeV}^{-1}   
&  {\rm for}\; x_{gq}^{}  \;=\;  0.3 
\vspace{1ex} \\   \displaystyle   
11\times 10^{-7}\;{\rm GeV}^{-1}   
&  {\rm for}\; x_{gq}^{}  \;=\;  1 
\vspace{1ex} \\   \displaystyle   
5\times 10^{-7}\;{\rm GeV}^{-1}   
&  {\rm for}\; x_{gq}^{}  \;=\;  4 
\end{array}      
\right.  
\;.   
\end{eqnarray}      
These limits exceed the SM values in  Eq.~(\ref{csm})  by up to 
several times.   
The resulting branching ratio can be as large as 
\begin{eqnarray}   
{\rm BR} \bigl( \Omega^-\rightarrow\Xi^-\gamma \bigr) _{\rm susy}  
\;\approx\;  5\times 10^{-5}   \;,     
\end{eqnarray}      
which is about an order of magnitude below the experimental limit.

In view of the potentially large values of  
$C_{\gamma,\rm susy}^{}$,  we should consider how they might affect 
the decay  $\,\Xi^-\rightarrow\Sigma^-\gamma,\,$  which has 
a measured branching-ratio of  $\,(1.27\pm 0.23)\times 10^{-4}.\,$   
This is because both  $\,\Xi^-\rightarrow\Sigma^-\gamma\,$  and  
$\,\Omega^-\rightarrow\Xi^-\gamma\,$  proceed by the same mechanisms 
at the quark level, due to the valence-quark content of the baryons 
involved~\cite{singer,GilWis}.  
To estimate the impact of the short-distance  
$\,s\rightarrow d\gamma\,$  operators on  
$\,\Xi^-\rightarrow\Sigma^-\gamma,\,$  
we again employ a chiral-Lagrangian analysis.  
Thus, we can construct  
\begin{eqnarray}     
{\cal L}_{BB'\gamma}^{}  \;=\;  
\tr \Bigl[ \bar{B}\, 
\bigl( \beta_+^{}+\beta_-^{}\gamma_5^{} \bigr) \sigma^{\mu\nu}\, 
\Bigl( \xi^\dagger \ell_{\mu\nu}^{} h \xi^\dagger  
      + \xi r_{\mu\nu}^{} h \xi \Bigr) \, 
B \Bigr]     
\;+\;  {\rm h.c.}   \;,      
\end{eqnarray}      
where  $\,\beta_\pm^{}=C_\gamma^\pm/(4\pi)^2.\,$  
As a result, we find that supersymmetric models can yield 
contributions to the branching ratio as much as 
$\,{\rm BR} \bigl( \Xi^-\rightarrow\Sigma^-\gamma \bigr) _{\rm susy}  
\approx 0.19\times 10^{-4}.\,$   
This is still consistent with the expectation that the decay is 
dominated by long-distance   
physics~\cite{singer,KogShi,EilIMS,singer2}.

\section{Conclusions}

We have investigated the effects of new physics beyond the standard 
model on the short-distance  $\,s\rightarrow d\gamma\,$  transition 
in  $\,\Omega^-\rightarrow\Xi^-\gamma\,$  decay.  
We have considered the possibility that the new interactions do not 
generate the chirality suppression occurring in the standard-model 
contribution to  $\,s\rightarrow d\gamma,\,$  thereby increasing 
the effect of the coupling by factors of a heavy-mass scale relative 
to the  $s$-quark mass.   
We have looked at contributions from two popular classes of models 
for new physics, left-right symmetric models and generic 
supersymmetric models.   
After taking the constraints from  $\,b\rightarrow s\gamma\,$  and   
kaon processes into account, we have found that contributions from   
the new interactions can be larger than that of the standard model   
by up to several times.    
The resulting effect on the branching ratio of 
$\,\Omega^-\rightarrow\Xi^-\gamma\,$  can be as large as an order 
of magnitude below the current experimental bound, which suggests 
that  $\,\Omega^-\rightarrow\Xi^-\gamma\,$  decay can be used 
to search for new physics or place constraints on it.   
This should encourage serious efforts to measure the decay.

\bigskip
\bigskip
  
{\bf Acknowledgments}$\;$    
I would like to thank G.~Valencia for helpful discussions.  
This work  was supported in part by DOE under contract number 
DE-FG02-92ER40730.


\begin{thebibliography}{99}
%
\bibitem{singer}
P. Singer, hep-ph/9607429, and references therein; 
L.~Bergstr\"{o}m and P.~Singer, Phys. Lett.~{\bf 169B}, 297~(1986).  
%  
\bibitem{KogShi}  
Ya.I. Kogan and M.A.~Shifman, 
Sov.~J. Nucl. Phys.~{\bf 38}, 628~(1983).  
%   
\bibitem{EilIMS}  
G.~Eilam, A.~Ioannissian, R.R.~Mendel, and P.~Singer, 
Phys. Rev.~D~{\bf 53}, 3629~(1996).
%   
\bibitem{NieBER}    
M.~Nielsen, L.A.~Barreiro, C.O.~Escobar, and R.~Rosenfeld, 
Phys. Rev.~D~{\bf 53}, 3620~(1996).    
%    
\bibitem{RafSin}   
R.~Safadi and P.~Singer, Phys. Rev.~D~{\bf 37}, 697~(1988).  
%     
\bibitem{EilIS}    
G.~Eilam, A.~Ioannissyan, and P.~Singer,   
Mod. Phys. Lett.~A~{\bf 11}, 2091~(1996).    
%  
\bibitem{pdb}   
C.~Caso et. al., Review of Particle Physics, 
Eur. Phys. J.~C~{\bf 3}, 1~(1998).
%
\bibitem{bsg1}
See, for example, J.A.~Hewett hep-ph/9406302; 
A.~Kagan and M.~Neubert,  Eur. Phys. J.~C~{\bf 7}, 5~(1999), 
and references therein.  
%   
\bibitem{dsg}  
X.-G.~He and G.~Valencia, Phys. Rev.~D~{\bf 61}, 075003 (2000); 
G.~Colangelo, G.~Isidori, and J.~Portoles, 
Phys. Lett.~B~{\bf 470}, 134~(1999);  
J.~Tandean and G.~Valencia, in preparation.  
%   
\bibitem{RosBEN}  
R.~Rosenfeld, L.A.~Barreiro, C.O.~Escobar, and M.~Nielsen, 
Phys. Rev.~D~{\bf 54}, 3645 (1996).
%  
\bibitem{mohapatra} 
R.N.~Mohapatra, {\it Unification and Supersymmetry} 
(Springer, New York, 1986), and references therein.  
%   
\bibitem{FujYam}
K.~Fujikawa and A.~Yamada, Phys. Rev.~D~{\bf 49}, 5890~(1994).
%   
\bibitem{bsg2}
P.~Cho and M.~Misiak, Phys. Rev.~D~{\bf 49}, 5894~(1994);
K.S.~Babu, K.~Fujikawa, and A.~Yamada, Phys. Lett.~B~{\bf 333}, 
196~(1994);  
T.G.~Rizzo, Phys. Rev.~D~{\bf 50}, 3303 (1994);    
C.-D.~L\"{u}, J.-L.~Hu, C.~Gao, {\it ibid}.~{\bf 52}, 4019~(1995). 
%   
\bibitem{GabGMS}   
F.~Gabbiani, E.~Gabrielli, A.~Masiero, and L.~Silvestrini, 
Nucl. Phys.~{\bf B477}, 321~(1996).   
%    
\bibitem{BurCIRS}
A.J.~Buras et al., Nucl. Phys.~B~{\bf 566}, 3~(2000).  
%    
\bibitem{nda}  
A.~Manohar and H.~Georgi, Nucl. Phys.~B~{\bf 234}, 189~(1984);  
H.~Georgi and L.~Randall, {\it ibid}.~{\bf 276}, 241~(1986).     
%    
\bibitem{AbdTV}  
A.~Abd~El-Hady, J.~Tandean, and G.~Valencia, 
Nucl. Phys.~{\bf A651}, 71~(1999).  
%     
\bibitem{InaLim}  
T.~Inami and C.S.~Lim, Prog. Theor. Phys.~{\bf 65}, 297~(1981); 
{\bf 65}, 1772(E)~(1981).  
%   
\bibitem{BerFG}   
S.~Bertolini, M.~Fabbrichesi, and E.~Gabrielli, 
Phys. Lett.~B~{\bf 327}, 136~(1994).  
%   
\bibitem{ShiVZ}   
M.~Shifman, A.~Vainshtein and V.~Zakharov, Phys. Rev.~D~{\bf 18},
2583~(1978); {\bf 19}, 2815(E)~(1979).
%    
\bibitem{BucBL}     
G.~Buchalla, A.J.~Buras, and M.E.~Lautenbacher, Rev. Mod. 
Phys.~{\bf 68}, 1125~(1996).  
%   
\bibitem{mele}
S.~Mele, Phys. Rev.~D~{\bf 59}, 113011~(1999).  
%  
\bibitem{PecZha} 
R.~Peccei and X.~Zhang, Nucl. Phys.~{\bf B337}, 269~(1990).  
%  
\bibitem{AbdVal}  
A.~Abd~El-Hady and G.~Valencia, Phys. Lett.~B~{\bf 414}, 173~(1997).  
%   
\bibitem{running}
B.~Grinstein, R.~Springer and M.~Wise, Phys. Lett.~B~{\bf 202}, 
138~(1988);   
R.~Grigjanis, P.J.~O'Donnell, M.~Sutherland, and H.~Navelet, 
{\it ibid}.~{\bf 213}, 355~(1988);  
G.~Cella, G.~Curci, G.~Ricciardi, and A.~Vicer\'{e},  
{\it ibid}.~{\bf 248}, 181~(1990);   
M.~Misiak, {\it ibid}.~{\bf 269}, 161~(1991);  
M. Ciuchini et al., {\it ibid}.~{\bf 316}, 127~(1993);   
Nucl. Phys.~{\bf B421}, 41~(1994).  
%  
\bibitem{MasMur}   
A.~Masiero and H.~Murayama, Phys. Rev. Lett.~{\bf 83}, 907~(1999).    
%   
\bibitem{GilWis}   
F.J.~Gilman and M.B.~Wise, Phys. Rev.~D.~{\bf 19}, 976~(1979).    
%  
\bibitem{singer2}   
P.~Singer, Phys. Rev.~D.~{\bf 42}, 3255~(1990).    
%   
\end{thebibliography}
\end{document}